\documentstyle[aps,preprint]{revtex}
\tightenlines

\begin{document}

\begin{titlepage}

\title{Exact solution of a Z(4) gauge Potts model on planar lattices}

\author{N.S.~Ananikian and R.G.~Ghulghazaryan\\[1mm]
{\small \sl Department of Theoretical Physics, 
Yerevan Physics Institute,} \\
{\small \sl Alikhanian Brothers 2, 375036 Yerevan, Armenia} \\[1mm]
N.Sh.~Izmailian${}^{*}$ \\[1mm]
{\small \sl Institute of Physics, Academia Sinica, Taipei 11529, Taiwan} \\[1mm]
R.~Shcherbakov\thanks{On leave of absence from Department of Theoretical Physics,
Yerevan Physics Institute, Armenia} \\[1mm]
{\small \sl Department of Geological Sciences,} \\
{\small \sl Snee Hall, Cornell University, Ithaca 14853, USA}}


\maketitle

\begin{abstract} 
The exact solution of a general Z(4) gauge Potts model with a single and 
double plaquette representation of the action is found on a subspace of 
gauge-coupling parameters  on the square 
and triangular lattices. The two Ising-type critical lines of a second-order 
phase transition for the model on a square lattice are found. For the model 
on a triangular lattice the two 
critical surfaces of an Ising-type and two nontrivial lines of a 
second-order phase transition with different critical behavior than on the
critical surfaces are found. It is shown that a two dimensional (2D) general 
Z(4) gauge Potts model 
with single and double plaquette representation of the action and a 2D 
spin-$\frac{3}{2}$ Ising model belong to the same universality class.\newline 

PACS number(s):05.50.+q, 64.60.Fr

\end{abstract}

\thispagestyle{empty}

\end{titlepage}

\section{Introduction}

Confinement-deconfinement phase transition in gauge theory is one of the 
challenging problems in modern physics. Lattice-gauge theory is a gauge-
invariant nonperturbative tool of regularization of gauge action to avoid 
the ultraviolet divergence in the theory. Although this regularization 
procedure is not unique, different ways of defining gauge theories on the 
lattice should lead to the same physics in each case when the continuum 
limit of vanishing lattice spacing is taken. From a theoretical point of view,
investigations with different lattice actions will enable a deeper understanding 
of the physics of confinement and other related problems in QCD.
The first lattice-gauge model with a single plaquette representation of the 
action has been introduced by Wilson,
\[
S=-\beta \sum_{p}{Re}\,U_{p}\,, 
\]
where $U_{p}$ denotes the usual plaquette variable, the product of link
gauge fields around a plaquette~\cite{wilson}. It was expected that non-Abelian 
gauge theories, in general, do not have any phase transitions separating strong- 
and weak-coupling regimes. Therefore, confinement, explicitly shown on the 
lattice in the strong-coupling region, should persist also in the continuum 
limit. Later, Bhanot and Creutz~\cite{bhanot1} extended the form of the Wilson 
action by adding an adjoint coupling term. Using Monte Carlo simulations it 
was shown that confinement could survive even through the phase diagram of 
the mixed action and the so-called bulk (volume) phase transitions separating 
strong- and weak-coupling regions exist~\cite{gavai,blum}.

The choice of action is still far from unique. Recently, several improved 
actions have been proposed as a way of reducing scaling violation in the 
approach to the continuum limit from a lattice action. Among them the 
Symanzik-Wiesz action constructed from a combination of ($1\times 1$) 
and ($1\times 2$) 
Wilson loops, the Bhanot-Creutz action~\cite{bhanot1}, the tadpole-improved 
actions~\cite{shakespeare}, and the $q$-state gauge Potts model with a single and 
double plaquette form of action~\cite{fanchiotti}.
  
The Monte Carlo analysis of the $q$-state Potts model with a single and 
double plaquette form of action~\cite{fanchiotti} showed that for $d=3$, 
$q=2$, first- and second-order transition lines; and for $d=2$, $q=3,4$, the 
second-order; and for $q=5$, first-order transitions are existed, which is 
in good agreement with the analytical results for $d=2$.

The lattice-gauge models with double plaquette interaction terms in the 
action were proposed and studied in three dimension and four dimension by Edgar~\cite{Edgar} and 
Bhanot{\it et al.}~\cite{Bhanot}. Turban investigated the two dimensional 
$2D$ gauge model with 
the global $Z(2)$ symmetry on a rectangular lattice~\cite{Turban}. 
He reduced it to the usual spin-$\frac{1}{2}$ Ising model on a
square lattice and obtained a point of a second-order phase transition. 
The $Z(3)$ gauge model on the flat triangular and square lattices with 
double plaquette representation of the action was investigated by Ananikian 
and Shcherbakov~\cite{Ananikian}. It was reduced to the spin-$1$
Blume-Emery-Griffiths (BEG) model~\cite{BEG} and an Ising-type critical 
line of a second-order phase transition was found on a subspace of the
interaction constants~\cite{Horig}.

The fact that lattice gauge theories could be mapped to the classical spin 
systems is well known. For example, Wilczek and Rajagopal~\cite{wilczek} 
showed that in QCD with two flavors of massless quarks, the chiral phase 
transition is in the same universality class as the classical O$(4)$ 
Heisenberg antiferromagnet and they also established a dictionary between 
QCD and the magnetic system (see also Ref.~\cite{iwasaki}). Okawa~\cite{okawa}, 
using Monte Carlo renormalization-group methods, showed that a $(3+1)$-dimensional 
SU$(2)$ lattice-gauge theory and a three-dimensional Ising model belong to the 
same universality class.

In this paper we consider the generalized $Z(4)$ gauge Potts model with a 
single and double plaquette representation of the action on the square and 
triangular lattices. We found an exact analytical solution of this model on a 
subspace of gauge-coupling parameters. Using duality transformation
~\cite{wannier} and exact results for the spin-$\frac{3}{2}$ Ising model on 
the square and honeycomb lattices~\cite{izmailian,ners}, we investigated the 
critical properties of the gauge theory. We showed that a two-dimensional 
generalized Z$(4)$ gauge Potts model with single and double plaquette 
representation of the action and a two-dimensional spin-$\frac{3}{2}$ Ising 
model belong to the same universality class.

The paper is organized as follows. In Sec.~II we define the model under 
consideration and present obtained results. Section~III contains some 
concluding remarks.

\section{The $Z(4)$ gauge Potts model}

The most general form of the $Z(4)$ gauge Potts model with a single and
double plaquette representation of the action is defined through the action 
\begin{equation}
S_{Gauge}=-\sum_{\langle p_{i}\,p_{j}\rangle}\sum_{y,z}\tilde{\beta}_{yz}\,\delta
_{U_{p_{i}},y}\,\delta _{U_{p_{j}},z}-\sum_{p_{i}}\sum_{z}\tilde{\beta}%
_{z}\,\delta _{U_{p_{i}},z}\,\,,  
\label{sg}
\end{equation}
where the outer summation in the first term runs over all nearest-neighbor
plaquettes and in the second one is over all plaquettes of the lattice. The
indexes $y$ and $z$ of the inner summations run over the group $Z(4)$. The 
$U_{p}=\prod_{b\in \partial p}U_{b}$ denotes the ordered product of link
gauge fields $U_{b}$'s around an elementary plaquette. Each link variable 
$U_{b}$ takes the value $\exp (ik\pi /2)\in Z(4)$, $k=0,1,2,3$. $\delta $ is
the standard Kronecker symbol and 
$\tilde{\beta}_{yz},\tilde{\beta}_{z},\,\,y,z\in Z(4)$ 
are coupling parameters.

From the obvious identity for the Kronecker symbols 
\[
\delta _{U_{p_{i}},1}+\delta _{U_{p_{i}},z_{1}}+\delta
_{U_{p_{i}},z_{2}}+\delta _{U_{p_{i}},z_{3}}=1\,, 
\]
where $z_{k}\in Z(4),\,k=1,2,3$, and an assumption that the coupling
parameters are symmetric under the transposition of the indexes $y$ and 
$z$, we can reduce the number of independent coupling parameters and 
rewrite gauge action~(\ref{sg}) in the following form 
\[
S_{Gauge}=-\sum_{\langle p_{i}\,p_{j}\rangle}\left[ \beta _{11}\delta
_{U_{p_{i}},1}\delta _{U_{p_{j}},1}+\beta _{22}\delta
_{U_{p_{i}},z_{1}}\delta _{U_{p_{j}},z_{1}}+\beta _{33}\delta
_{U_{p_{i}},z_{2}}\delta _{U_{p_{j}},z_{2}}+\right. 
\]
\begin{equation}
\left. \beta _{12}\left( \delta _{U_{p_{i}},1}\delta
_{U_{p_{j}},z_{1}}+\delta _{U_{p_{i}},z_{1}}\delta _{U_{p_{j}},1}\right)
+\beta _{13}\left( \delta _{U_{p_{i}},1}\delta _{U_{p_{j}},z_{2}}+\delta
_{U_{p_{i}},z_{2}}\delta _{U_{p_{j}},1}\right) +\right.  \label{sgred}
\end{equation}
\[
\left. \beta _{23}\left( \delta _{U_{p_{i}},z_{1}}\delta
_{U_{p_{j}},z_{2}}+\delta _{U_{p_{i}},z_{2}}\delta _{U_{p_{j}},z_{1}}\right)
\right] +\sum_{p_{i}}\left( \beta _{1}\delta _{U_{p_{i}},1}+\beta _{2}\delta
_{U_{p_{i}},z_{1}}+\beta _{3}\delta _{U_{p_{i}},z_{2}}\right) \,. 
\]
The partition function for this gauge model is defined as a sum of Boltzmann
weights $\exp (S_{Gauge})$ over all configurations of the gauge variables 
$\{U\}$,
\begin{equation}
Z_{Gauge}=\sum_{\{U\}}\exp \left[ -S_{Gauge}\right] \,.  \label{zgauge}
\end{equation}
To establish the connection between this gauge model and the spin-$\frac{3}{2}$ 
Ising model, we introduce spin variables $S_{i}$ in the sites of the dual
lattice such that

\begin{equation}
\begin{array}{lcl}
S_{i} & = & \frac{1}{2}(\delta _{U_{p_{i}},z_{0}}-\delta _{U_{p_{i}},z_{1}})+%
\frac{3}{2}(\delta _{U_{p_{i}},z_{2}}-\delta _{U_{p_{i}},z_{3}})\,, \\[4mm] 
S_{i}^{2} & = & \frac{1}{4}(\delta _{U_{p_{i}},z_{0}}+\delta
_{U_{p_{i}},z_{1}})+\frac{9}{4}(\delta _{U_{p_{i}},z_{2}}+\delta
_{U_{p_{i}},z_{3}})\,, \\[4mm] 
S_{i}^{3} & = & \frac{1}{8}(\delta _{U_{p_{i}},z_{0}}-\delta
_{U_{p_{i}},z_{1}})+\frac{27}{8}(\delta _{U_{p_{i}},z_{2}}-\delta
_{U_{p_{i}},z_{3}})\,.
\end{array}
\label{ss}
\end{equation}
Using this substitution, we can rewrite gauge action~(\ref{sgred}) in
terms of new spin variables 
\[
S_{Spin}=-\sum_{\langle ij\rangle}\left[
J\,S_{i}S_{j}+K\,S_{i}^{2}S_{j}^{2}+L\,S_{i}^{3}S_{j}^{3}+\frac{M}{2}%
\,(S_{i}S_{j}^{3}+S_{i}^{3}S_{j})+\right. 
\]
\begin{equation}
\left. \frac{M_{1}}{2}\,(S_{i}S_{j}^{2}+S_{i}^{2}S_{j})+\frac{M_{2}}{2}%
\,(S_{i}^{2}S_{j}^{3}+S_{i}^{3}S_{j}^{2})\right] -\sum_{i}\left(
h\,S_{i}-\Delta \,S_{i}^{2}+h_{3}\,S_{i}^{3}\right)\,,  
\label{sspin}
\end{equation}
where constants $J$, $K$, $L$, $M$, $M_{1}$, $M_{2}$, $h$, $\Delta$ and $h_{3}$ 
are linear combinations of the original gauge-coupling parameters: 
\begin{eqnarray}
J &=&\frac{81}{64}\beta _{11}-\frac{81}{32}\beta _{12}-\frac{3}{32}\beta
_{13}+\frac{81}{64}\beta _{22}+\frac{3}{32}\beta _{23}+\frac{1}{576}\beta
_{33}, \medskip \nonumber \\
K &=&\frac{1}{16}\beta _{11}+\frac{1}{8}\beta _{12}-\frac{1}{8}\beta _{13}+%
\frac{1}{16}\beta _{22}-\frac{1}{8}\beta _{23}+\frac{1}{16}\beta _{33}, 
\medskip \nonumber \\
L &=&\frac{1}{4}\beta _{11}-\frac{1}{2}\beta _{12}-\frac{1}{6}\beta _{13}+%
\frac{1}{4}\beta _{22}+\frac{1}{6}\beta _{23}+\frac{1}{36}\beta _{33},
\medskip \nonumber \\
M &=&-\frac{9}{8}\beta _{11}+\frac{9}{4}\beta _{12}+\frac{5}{12}\beta _{13}-%
\frac{9}{8}\beta _{22}-\frac{5}{12}\beta _{23}-\frac{1}{72}\beta _{33}, 
\medskip \nonumber \\
M_{1} &=&-\frac{9}{16}\beta _{11}+\frac{7}{12}\beta _{13}+\frac{9}{16}\beta
_{22}-\frac{13}{24}\beta _{23}-\frac{1}{48}\beta _{33}, 
\medskip \nonumber \\
M_{2} &=&\frac{1}{4}\beta _{11}-\frac{1}{3}\beta _{13}-\frac{1}{4}\beta
_{22}+\frac{1}{6}\beta _{23}+\frac{1}{12}\beta _{33}, 
\medskip \nonumber \\
h &=&\gamma \left( \frac{81}{128}\beta _{11}-\frac{3}{32}\beta _{13}-\frac{81%
}{128}\beta _{22}+\frac{3}{64}\beta _{23}+\frac{1}{384}\beta _{33}\right) +%
\frac{9}{8}\beta _{1}-\frac{9}{8}\beta _{2}-\frac{1}{24}\beta _{3}, 
\medskip \nonumber \\
\Delta &=&\gamma \left( \frac{9}{64}\beta _{11}+\frac{9}{32}\beta _{12}-%
\frac{5}{32}\beta _{13}+\frac{9}{64}\beta _{22}-\frac{5}{32}\beta _{23}+%
\frac{1}{64}\beta _{33}\right) +\frac{1}{4}\beta _{1}+\frac{1}{4}\beta _{2}-%
\frac{1}{4}\beta _{3}, 
\medskip \nonumber \\
h_{3} &=&\gamma \left( -\frac{9}{32}\beta _{11}+\frac{1}{8}\beta _{13}+\frac{%
9}{32}\beta _{22}+\frac{1}{16}\beta _{23}-\frac{1}{96}\beta _{33}\right) -%
\frac{1}{2}\beta _{1}+\frac{1}{2}\beta _{2}+\frac{1}{6}\beta _{3}\,,
\nonumber
\end{eqnarray}
where $\gamma$ is the coordination number of the dual lattice. 
Action~(\ref{sspin}) coincides with the Hamiltonian multiplied by 
$1/k_{B}T$ of the spin-$\frac{3}{2}$ Ising model~\cite{izmailian,ners}. 
Thus, the partition function of generalized gauge Potts 
model~(\ref{zgauge}) is equal up to the factor to the partition function 
of the spin-$\frac{3}{2}$ Ising model defined on the dual lattice 
\begin{equation}
Z_{Gauge}=4^{a(\gamma)N}Z_{Spin}^{Dual}\,,  \label{zgt}
\end{equation}
where 
\[
Z_{Spin}^{Dual}=\sum_{\{S\}}\exp \left[ -S_{Spin}\right] \,, 
\]
and $a(\gamma)$ is a constant that depends on the coordination number 
of the lattice, $a=1$ for the square lattice and $a=\frac{1}{2}$ for the honeycomb 
one respectively.

A factor $4^{a(\gamma)N}$ has been included in Eq.~(\ref{zgt}) to
take into account the difference between the number of gauge $\{U\}$ and
spin $\{S\}$ configurations. To obtain the phase structure of this gauge
model we will restrict ourselves to the spin-$\frac{3}{2}$ Ising model on
the square and honeycomb lattices. For coincidence with the 
spin-$\frac{3}{2}$ Ising model the coefficients $h$, $h_{3}$, $M_{1}$ and  
$M_{2}$ will be set to zero. Thus, we can express the parameters of 
the spin-$\frac{3}{2}$ model through the rest of the gauge couplings as follows: 
\begin{eqnarray}
J &=&\frac{337}{288}\beta _{11}-\frac{81}{32}\beta _{12}+\frac{1}{288}\beta
_{13}+\frac{49}{36}\beta _{22},  \label{J} \nonumber \\
K &=&\frac{1}{8}\beta _{11}+\frac{1}{8}\beta _{12}-\frac{1}{8}\beta _{13},%
  \label{K} \nonumber \\
L &=&\frac{1}{18}\beta _{11}-\frac{1}{2}\beta _{12}+\frac{1}{18}\beta _{13}+%
\frac{4}{9}\beta _{22},  \label{L} \nonumber \\
M &=&-\frac{25}{36}\beta _{11}+\frac{9}{4}\beta _{12}-\frac{1}{36}\beta
_{13}-\frac{14}{9}\beta _{22},  \label{M} \nonumber \\
\Delta &=&\gamma \left( \frac{9}{32}\beta _{11}+\frac{9}{32}\beta _{12}-%
\frac{1}{32}\beta _{13}\right) +\frac{1}{2}\beta _{1}. \nonumber  
\end{eqnarray}

The general spin-$\frac{3}{2}$ Ising model on a honeycomb lattice 
was investigated by Izmailian and Ananikian~\cite{ners}. 
The model is described by the Hamiltonian 
\[
-\beta H=\sum_{\langle ij\rangle}\left[
J\,S_{i}S_{j}+K\,S_{i}^{2}S_{j}^{2}+L\,S_{i}^{3}S_{j}^{3}+\frac{M}{2}%
\,(S_{i}S_{j}^{3}+S_{i}^{3}S_{j})\right] -\Delta \,\sum_{i}S_{i}^{2}\,,
\]
where $\beta=1/k_{B}T$, 
$S_{i}=\pm \frac{1}{2},$ $\pm \frac{3}{2}$ is the spin variable at 
site $i$, and $\langle ij\rangle$ indicates the summation over all nearest-neighbor pairs of
sites. Under the conditions 
\begin{equation}
\begin{array}{rcl}
\tanh^{2}(J_{1}) &=&\tanh(J_{2})\tanh(J_{0}),  \\
\exp \left( -4K\right) &=&\cosh \left( J_{2}-J_{0}\right)\,,  
\label{cond}
\end{array}
\end{equation}
that in terms of gauge-coupling parameters are
\begin{equation}
\begin{array}{rcl}
J_{0}&=&\frac{1}{4}\beta_{11}-\frac{1}{2}\beta_{12}+\frac{1}{4}\beta_{22},\\
J_{1}&=&\frac{1}{4}\beta_{11}-\frac{1}{4}\beta_{22},\\
J_{2}&=&-\frac{1}{4}\beta_{11}+\frac{1}{2}\beta_{13}+\frac{1}{4}\beta_{22}\,,  
\label{cond1}
\end{array}
\end{equation}
the model transforms to the spin-$\frac{1}{2}$ Ising model on the same
lattice. The free energy and critical point for the spin-$\frac{1}{2}$ Ising
model on the honeycomb lattice in the limit of an infinite lattice are well
known~\cite{syozi}. Thus, using this result one can obtain the important
thermodynamic properties of the spin-$\frac{3}{2}$ Ising model with $Z(2)$
symmetry~\cite{izmailian}. After substitution of $J$, $K$ and $L$ from 
Eq.~(\ref{cond1}) into Eq.~(\ref{cond}) we obtain the subspace in which the corresponding 
spin-$\frac{3}{2}$ Ising model can be solved exactly, 
\begin{eqnarray}
2-\exp(\beta _{12})-\exp \left( \beta _{11}-\beta
_{13}\right) &=&0,  \nonumber \\
\cosh \frac{1}{2}\left( \beta _{11}-\beta _{22}\right) \cosh \frac{1}{2}%
\left( \beta _{11}-\beta _{13}-\beta _{12}\right) &=&\cosh \frac{1}{2}\left(
\beta _{22}+\beta _{13}-\beta_{12}\right)\,.  
\label{horcond}
\end{eqnarray}
Then, using the exact solution~\cite{ners}, we obtain the 
$\lambda $ surface of an Ising-type transition 
(logarithmic specific heat singularity) for our Z$(4)$ gauge model 
\begin{equation}
\frac{\tanh \frac{1}{4}\left( \beta _{22}-2\beta _{12}+\beta _{11}\right)
+\tanh \frac{1}{4}\left( \beta _{22}+2\beta _{13}-\beta _{11}\right) \exp
\left( -2\Delta _{0}\right) }{1+\exp \left( -2\Delta _{0}\right) }=\frac{1}{%
\sqrt{3}}\,,  
\label{crit}
\end{equation}
where 
\begin{eqnarray*}
\Delta _{0} &=&\frac{3}{32}\left( 9\beta _{11}+9\beta _{12}-\beta
_{13}\right) +\frac{1}{2}\beta _{1}-3R, \\
\exp (-4R) &=&\frac{\cosh \frac{1}{4}\left( \beta _{22}-2\beta _{12}+\beta
_{11}\right) }{\cosh \frac{1}{4}\left( \beta _{22}+2\beta _{13}-\beta
_{11}\right) }\cosh ^{\frac{5}{4}}\frac{1}{2}\left( \beta _{12}+\beta
_{13}-\beta _{11}\right),
\end{eqnarray*}
in the space spanned by $\beta_{12}$, $\beta_{13}$, and $\beta_{1}$. It
is easy to see that the $\lambda $ surface of the critical points in 
Eq.~(\ref{crit}) is defined only in the two regions of the ($\beta _{12}$, 
$\beta_{13}$) plane,

\begin{itemize}
\item[(i)]  $0\leq \tanh \frac{1}{4}(\beta _{22}-2\beta _{12}+\beta
_{11})\leq 1/\sqrt{3}$ \hspace{5mm} and \\[2mm]
$1/\sqrt{3}\leq \tanh \frac{1}{4}\left( \beta
_{22}+2\beta _{13}-\beta _{11}\right) \leq 1$,

\item[(ii)]  $0\leq \tanh \frac{1}{4}(\beta _{22}+2\beta _{13}-\beta
_{11})\leq 1/\sqrt{3}$ \hspace{5mm} and \\[2mm]
$1/\sqrt{3}\leq \tanh \frac{1}{4}\left( \beta
_{22}-2\beta _{12}+\beta _{11}\right) \leq 1$.
\end{itemize}

For each set of $\beta _{12}$ and $\beta _{13}$, Eq.~(\ref{crit}) determines
the unique value of $\Delta $, except for the intersecting point of the two 
regions (i) and (ii) for which $\beta _{1}$ is an arbitrary. Thus, the 
$\lambda $ surface in Eq.~(\ref{crit}) contains two nontrivial 
$\lambda $-lines of critical points given by

\begin{itemize}
\item[(a)]  $\beta _{11}=\beta _{13}=2\ln (2+\sqrt{3}),\quad \beta
_{12}=\beta _{22}=0,\quad \beta _{1}$-arbitrary,

\item[(b)]  $\beta _{11}=\beta _{12}=\beta _{13}=0,\quad \beta
_{22}=2\ln (2+\sqrt{3}),\quad \beta_{1}$-arbitrary.
\end{itemize}

As shown in \cite{ners} on the $\lambda$ lines the model exhibits a critical behavior different
from the critical behavior elsewhere on the $\lambda $ surface. The phase
transition is not marked with the logarithmic divergence of the derivative
 of the order parameter $P$, where $P$ is the quadrupolar moment defined as
\[
P=\frac{1}{N}\sum_{i}^{N}\left\langle S_{i}^{2}\right\rangle
=Z^{-1}\sum_{\left\{ S\right\} }S_{i}^{2}\exp \left( -\beta H\right)\,. 
\]
This phase transition is associated with the logarithmic divergence in the
specific heat.

The area in the plane of coupling parameters $\exp(\beta _{12})$ and 
$\exp(\beta_{13})$, 
where the $\lambda $ surface exists is shown in Fig.~1. In the Appendix it
is proved that this area is connected and 
 there is no phase transition for $T\rightarrow \infty $ and $%
T\rightarrow 0$. Hence, for all possible values of the coupling
parameters there is only one finite critical value of the external field 
for which the phase transition is of the second-order 
except points (a) and (b) in the above equation.

The spin-$\frac{3}{2}$ Ising model with $Z(2)$ symmetry was investigated on a
square lattice by Izmailian~\cite{izmailian}. It was shown that this model
is reducible to an eight-vertex model on a surface in the parameter space
spanned by the coupling constants $J$, $K$, $L$, and $M$. It was also shown that this
model is equivalent to an exactly solvable free fermion model along two
lines in the parameter space. The two $\lambda $ lines of a second-order
phase transition was found exactly in this model.

In terms of our gauge theory, these $\lambda $ lines have the following form:

\begin{itemize}
\item[(c)]  $\beta _{11}=\beta _{12}=\beta_{13}=0,\quad \beta
_{22}=2\ln (1+\sqrt{2}),\quad \beta _{1}$-arbitrary,

\item[(d)]  $\beta _{11}=\beta _{13}=2\ln (1+\sqrt{2}),\quad
\beta _{12}=\beta _{22}=0,\quad \beta _{1}$-arbitrary.
\end{itemize}
On these $\lambda $ lines our gauge theory exhibits an Ising-type
second-order phase transition (logarithmic specific-heat singularity).

Thus, we showed that there exists an area (lines for square lattice) where the 
$2$D generalized Z$(4)$ Potts gauge model mapped to the corresponding $2$D 
spin-$\frac{3}{2}$ Ising model. Hence, because of universality of critical 
indexes it follows that these two models have the same critical indexes and 
belong to the same universality class. 
\section{Concluding remarks}

In summary, we have found an exact analytical solution of the Z$(4)$ 
gauge-lattice model with a single and double plaquette representation of the 
action by mapping it to the dual spin-$\frac{3}{2}$ Ising model with $Z(2)$ 
symmetry. For the model on the square lattice we found the $\lambda$ lines 
of the second-order phase transition with logarithmic specific-heat 
singularity. For the model on the triangular lattice we derived the two 
$\lambda $ surfaces of a second-order phase transition with a usual 
Ising-type singularity of the order parameter and two nontrivial 
$\lambda $ lines of critical points on which our model exhibits 
the critical behavior unlike critical behavior elsewhere 
on the $\lambda $ surfaces. We demonstrated that the $2$D general Potts gauge 
model belongs to the same universality class as the $2$D spin-$\frac{3}{2}$ Ising 
model.

\section*{Acknowledgments}

We would like to thank R.~Flume for fruitful discussions. R.S. is indebted
to V.B.~Priezzhev and D.L.~Turcotte for stimulating discussions. This work
was partially supported by Grant Nos. INTAS-96-690 and INTAS-97-347, and ISTC
Project No. A-102.

\section*{Appendix}

Here we present the analytic investigation of the $\lambda$ surface in the
plane of the coupling parameters $\exp(\beta_{12})$ and $\exp(\beta_{13})$
for triangular lattice. Let us make the following denotations 
$x=\exp(\beta_{12})$, $y=\exp(\beta_{13})$ and 
$z=\exp(\beta_{22})$. After elimination of $\beta_{11}$ from 
Eq.~(\ref{horcond}) one obtains the first-order polynomial for $z$,
$$
z(1-2y+xy)+(2-x)(y-x)=0. \eqno (A1)
$$
In terms of the variables $x$, $y$ and $z$ the conditions (i) and (ii) 
take the following form:
\begin{itemize}
\item[A(i)]  $1\leq zy(2-x)/x^2\leq c^2 \;$ and $\; zy/(2-x)\geq c^2,$

\item[A(ii)]  $1\leq zy/(2-x)\leq c^2 \;$ and $\; zy(2-x)/x^2\geq c^2$,
\end{itemize}
where $c=2+\sqrt{3}$. From A(i) and A(ii) it is easy to show that $x<2$ 
for any values of $z$ and $y$. Taking into account that for $T\rightarrow 0 \;$ 
$x,y,z$ can take only values \{$0,1,\infty$\} and for $T\rightarrow \infty \;$ 
$x=y=z=1$, one can show that neither A(i) nor A(ii) are satisfied, hence 
there is no phase transition for $T\rightarrow 0$ and 
$T\rightarrow \infty $. 
 Using this fact one can construct the area in the plane of 
coupling parameters $\exp(\beta _{12})$ and $\exp(\beta _{13})$  
in which Eq.~(A1) and one of the conditions (Ai) or (Aii) are 
satisfied. This area is shown in Fig.~1.

\pagebreak
\begin{center}
FIGURE CAPTION
\end{center}
\vspace{2cm} 
Fig 1. The area in the plane of the coupling parameters
$\exp{\beta_{12}}$ and $\exp{\beta_{13}}$ where the
critical $\lambda$ surface exists, i.e., conditions (\ref{horcond}) 
and (i) or (ii) are satisfied. Points A and B correspond to projections of
 two nontrivial $\lambda$ lines of (a) critical points and (b) for which 
 $\beta_1$ is arbitrary.

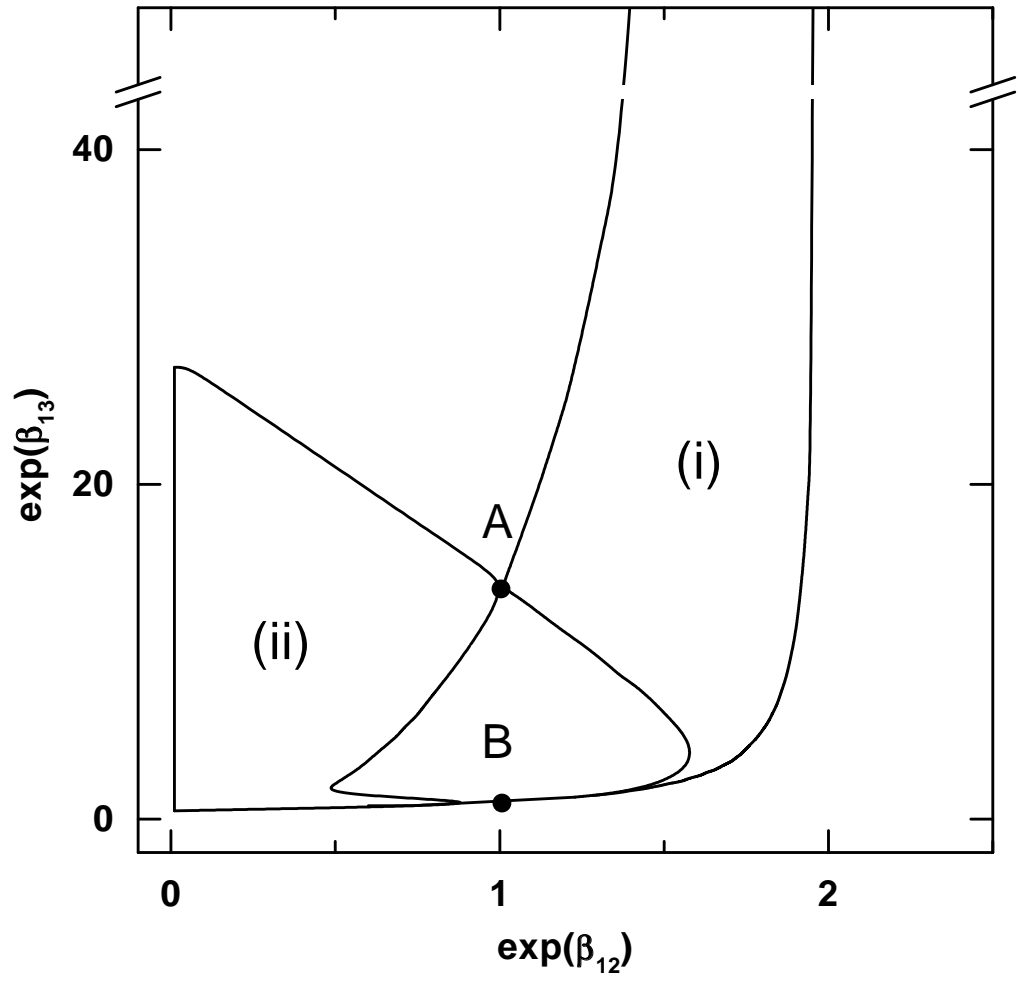
\end{document}